\documentclass [twocolumn,aps,prb]{revtex4}
\usepackage{epsfig,graphicx,amssymb,color,amssymb,amsmath}

\begin{document}
\title{Probing Plasmons in Graphene by Resonance Energy Transfer}

\author{Kirill A. Velizhanin}\email{kirill@lanl.gov}
\affiliation{Center for Nonlinear Studies (CNLS)/T-4, Theoretical Division, Los Alamos National Laboratory, Los Alamos, NM 87545 }
\author{Anatoly Efimov}
\affiliation{Center for Integrated Nanotechnologies, Materials Physics \& Applications Division, Los Alamos National Laboratory, Los Alamos, NM 87545}

\date{\today}

\begin{abstract}
We theoretically propose an experimental method to probe electronic excitations in graphene -- monoatomic layer of carbon -- by monitoring the fluorescence quenching of a semiconductor quantum dot (or a dye molecule) due to the resonance energy transfer to the graphene sheet. We show how the dispersion relation of plasmons in graphene can be accurately extracted by varying the back-gate voltage and the distance between the quantum dot and graphene.
\end{abstract}

\maketitle


\section{Introduction}

Many appealing properties of graphene -- a monoatomic crystalline sheet of carbon -- stem from its unique electronic structure.\cite{Geim2007-183} Specifically, its honeycomb lattice combined with the conjugation of $\pi$-electrons over the entire sheet results in the electronic spectrum of a zero-gap semiconductor with ``ultra-relativistic'' electrons and holes.\cite{Wallace1947-622} Already this makes graphene enormously appealing from both basic and application standpoints. What makes graphene even more attractive is the possibility to tune these properties in a wide range by patterning, chemical functionalization, doping etc. For example, shifting the Fermi level away from the charge neutrality (Dirac) point by applying the back-gate voltage, and, therefore, changing the graphene's electrical conductivity, can become useful in graphene-based electronics.\cite{Berger2004-19912,Geim2007-183}

The improved electrical conductivity of a back-gated or chemically doped graphene sample leads to qualitative changes in its optical properties. In particular, collective excitations (plasmons), rather then single-particle excitations (electron-hole pairs), are expected to define an electronic response of graphene to a low-frequency optical perturbation. Plasmons in graphene are of great interest from the basic perspective, as collective excitations in a two-dimensional electron gas with very peculiar properties. Besides, the graphene plasmonics holds promise for, e.g., photonic and optoelectronic devices, as well as metamaterials.\cite{Bonaccorso2010-611,Papasimakis2010-8353} Plasmons in graphene have been predicted and studied, theoretically\cite{Shung1986-979,Wunsch2006-318,Hwang2007-205418,Polini2008-081411,Hill2009-27005,Jablan2009-245435,Muniz2010-081411} but experimental studies of this phenomenon are still very sparse.\cite{Liu2008-201403,Eberlein2008-233406,Tegenkamp2011-012001}

In this paper we propose an experimental technique to probe plasmons in graphene. The technique is based on the F\"orster resonance energy transfer between a fluorescent semiconductor quantum dot (or a dye molecule) and a nearby graphene layer. Efficient energy transfer between a dye molecule and graphene has been theoretically predicted by Swathi and Sebastian\cite{Swathi2008-054703,Swathi2009-086101,Swathi2009-777} and later confirmed experimentally.\cite{Chen2010-2964} However, Swathi and Sebastian described electronic excitations in graphene on the single-particle level, which is accurate only for a nearly undoped (charge-neutral) graphene. In this paper, electronic excitations are treated more accurately by adopting the random-phase approximation, which allows for recovering the collective electronic behavior. We demonstrate that electronic excitations in graphene, both single-particle and collective, can be sensitively probed by studying the fluorescence quenching of the quantum dot due to the energy transfer to graphene. Specifically, we show how the plasmon dispersion can be extracted from experiment. 

The expected advantage of the proposed technique over the typically used electron energy loss spectroscopy (EELS)\cite{Eberlein2008-233406,Liu2008-201403,Tegenkamp2011-012001} is its intrinsic locality, i.e., plasmons are proposed to be probed {\em locally} by a semiconductor quantum dot (typically a few nanometers in diameter). In contrast, EELS averages the plasmonic response over a certain portion of a graphene sample, thus, adding the inhomogeneous broadening due to, e.g., charge puddles,\cite{Martin2008-144} to experimental observables.

The paper is organized as follows. The general theory of quantum dot fluorescence quenching due to the resonance energy transfer to graphene is given in Sec.~\ref{sec:fqe}. The analysis of quenching efficiency within the single-particle and random-phase approximation levels is provided in Sec.~\ref{sec:plasmon} and Sec.~\ref{sec:sp}, respectively. Sec.~\ref{sec:conclusion} concludes. 

\section{F\"orster resonance energy transfer}
\label{sec:fqe}

F\"orster resonance energy transfer (FRET) refers to the transfer of electronic excitation energy between chromophores\cite{fn1} mediated by the nonradiative Coulomb coupling.\cite{Novotny2006} This process consists of deexcitation of an initially (optically) excited donor chromophore and the simultaneous excitation of an acceptor chromophore. An example of such a process is the energy transfer between a semiconductor quantum dot (QD) and a non-fluorescent organic molecule as an acceptor chromophore.\cite{Willard2003-575} As the energy transfer to the non-fluorescent ``dark'' molecule (i.e., quencher) competes with the intrinsic fluorescence of the ``bright'' QD, the FRET rate can be assessed through the effective decrease (quenching) of the QD fluorescence quantum yield, and/or through the shorter apparent fluorescence lifetime, measured by time-resolved fluorescence spectroscopy. Specifically, the apparent fluorescence lifetime is given by $\tau=1/(\tau^{-1}_0+k_q)$, where $\tau_0$ is the lifetime of the isolated QD and $k_q$ is the FRET (quenching) rate. Accordingly, the decreased fluorescence quantum yield in the presence of FRET is given by $\tau/\tau_0$. Therefore, $k_q$ can be extracted provided $\tau$ and $\tau_0$ are known from experiment.

The efficiency of FRET is strongly affected by parameters of the acceptor's excitation spectrum. A trivial example is a vanishing FRET rate in a system where an acceptor chromophore does not have excitations resonant to the lowest excited state of a donor chromophore. This allows one to use FRET as a spectroscopic tool to probe an electronic structure of a system of interest. An advantage of this method is the near-field regime of the donor-acceptor interaction, which allows for probing excitations forbidden in the far-field (optical) regime due to, e.g., the large wavelength mismatch between optical photons and material electronic excitations. Short-wavelength character of electronic excitations in graphene, especially that of plasmons,\cite{Shung1986-979,Wunsch2006-318,Hwang2007-205418} hinders the study of these excitations by standard far-field optical techniques. We propose to use FRET between QD and a graphene sheet to probe electronic excitations in graphene.  The theoretical framework for FRET in the QD-graphene complex is developed in the rest of the section.

\subsection{Fluorescence quenching efficiency}

We define the wavefunctions of the excited and the ground states of QD as $|e\rangle$ and $|g\rangle$, respectively. The true many-body ground state and excited states of graphene are denoted by $|0\rangle$ and $|n\rangle$, respectively. The quenching rate, $k_q$, for the energy transfer from QD to graphene is given by Fermi's golden rule as 
\begin{equation}
k_q=2\pi\hbar^{-1}\sum_n|\langle n|\langle g|\hat{V}|e\rangle|0\rangle|^2\delta(\epsilon-E_n),
\label{eq:qrate}
\end{equation}
where $\hat{V}$ is the operator of Coulomb interaction between fluctuating charge densities of QD and graphene
\begin{equation}
\hat{V}=\int d{\bf r}\:V({\bf r})\left(|e\rangle\langle g|+|g\rangle\langle e|\right)\hat{\rho}({\bf r}).
\end{equation}
The operator of graphene charge density is given by $-e\hat{\rho}({\bf r})=-e\hat{\varphi}^\dagger({\bf r})\hat{\varphi}({\bf r})$, where operators $\hat{\varphi}^\dagger({\bf r})$ and $\hat{\varphi}({\bf r})$ create and destroy an electron at position ${\bf r}$ within the graphene sheet, respectively. The absolute value of the electron charge is denoted by $e=|e|$. The excitation energies of QD and graphene are denoted by $\epsilon$ and $E_n$, respectively. Vector variables are denoted in bold. 

For the rest of the paper, we adopt the dipole approximation for QD, which results in $V({\bf r})=-e({\bf d}\cdot{\bf r})/r^3$. The transition dipole of QD is given by ${\bf d}$. This approximation is accurate if $z\gg D$, where $z$ is the distance between QD and graphene, and $D$ is the QD diameter. The validity of this approximation for the realistic case of PbSe QD is discussed in Sec.~\ref{sec:plasmon}.

The quenching rate in Eq.~(\ref{eq:qrate}) can be rewritten through the retarded polarization operator of graphene $\Pi^r(q,\epsilon)$ as (a detailed derivation is given in the Appendix)
\begin{equation}
k_q=-2\pi e^2\hbar^{-1}(d^2_\parallel+2d^2_\perp)\int^{\infty}_0 q dq\:
{\rm Im}\left[\Pi^r(q,\epsilon)\right]e^{-2qz},
\end{equation}
where $d_\parallel$ and $d_\perp$ are the projections of the QD transition dipole ${\bf d}$ onto the graphene's plane and the normal to this plane, respectively. If we ignore for simplicity that the population of the QD excited state depends on the angle between the QD transition dipole and the polarization of the initial QD-exciting laser pulse, we simply need to average over all the possible orientations of ${\bf d}$ with respect to the graphene's plane which yields 
$\langle d^2_\parallel\rangle/2=\langle d^2_\perp\rangle=d^2/3$ resulting in
\begin{equation}
k_q=-\frac{8\pi e^2 d^2}{3\hbar}\int^{\infty}_0 q dq\:
{\rm Im}\left[\Pi^r(q,\epsilon)\right]e^{-2qz}.
\label{eq:qratef}
\end{equation}
We define the quenching efficiency (QE) as $\varphi_q=k_q/k_r$, where $k_r$ is the fluorescence rate for the isolated QD\cite{Landau1982-v4}
\begin{equation}
k_r=\frac{4\epsilon^3}{3\hbar^4c^3}d^2.
\label{eq:SER}
\end{equation}
Substituting Eq.~(\ref{eq:SER}) into Eq.~(\ref{eq:qratef}), one obtains for QE
\begin{equation}
\varphi_q=-\frac{2\pi\hbar^3c^3e^2}{\epsilon^3}\int^\infty_0 q dq\:
{\rm Im}\left[\Pi^r(q,\epsilon)\right]e^{-2qz}
\label{eq:QE}.
\end{equation}

\subsection{Massless Dirac fermions approximation}

\begin{figure}
\centering{}
\includegraphics[width=7.5cm]{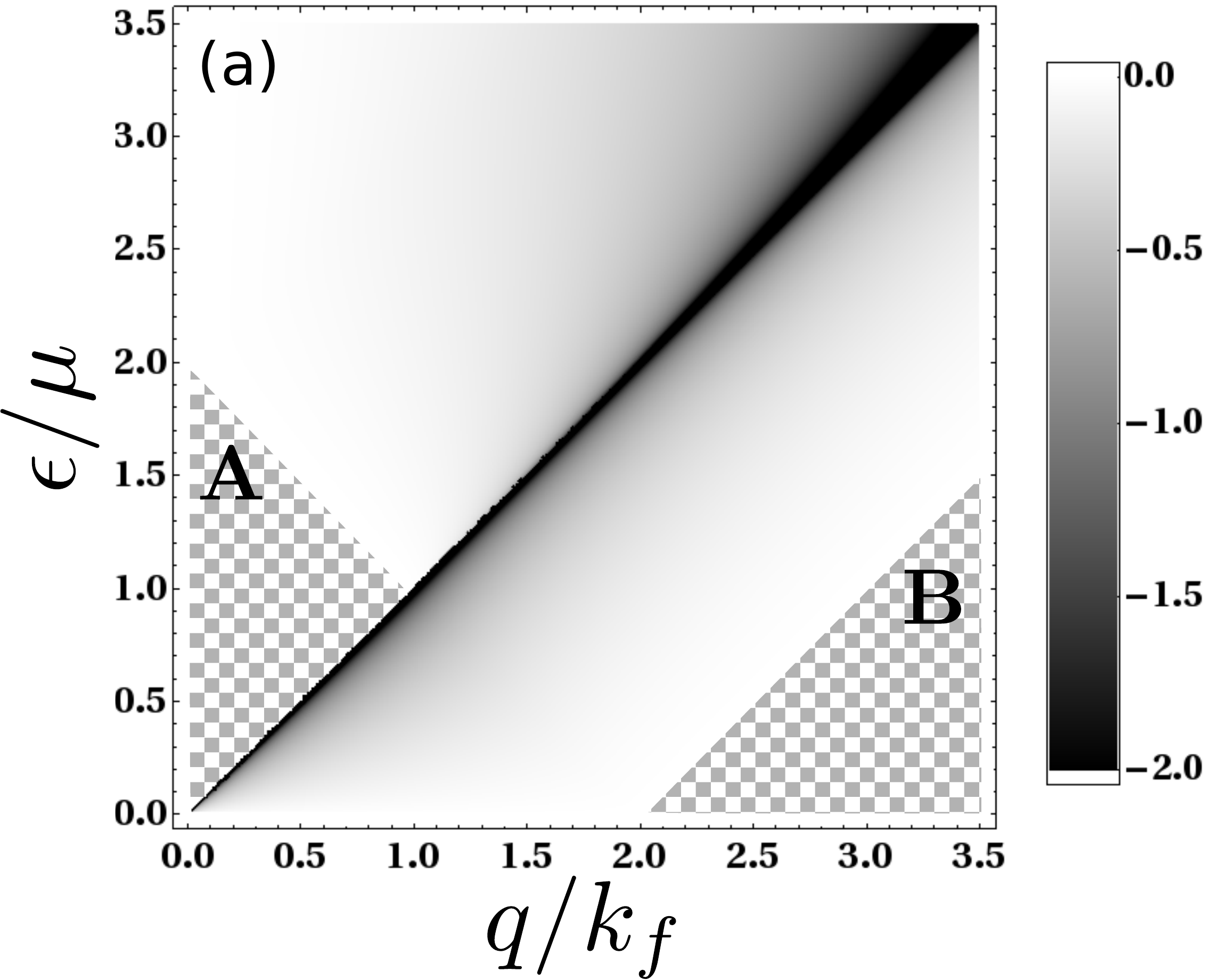}
\includegraphics[width=7.5cm]{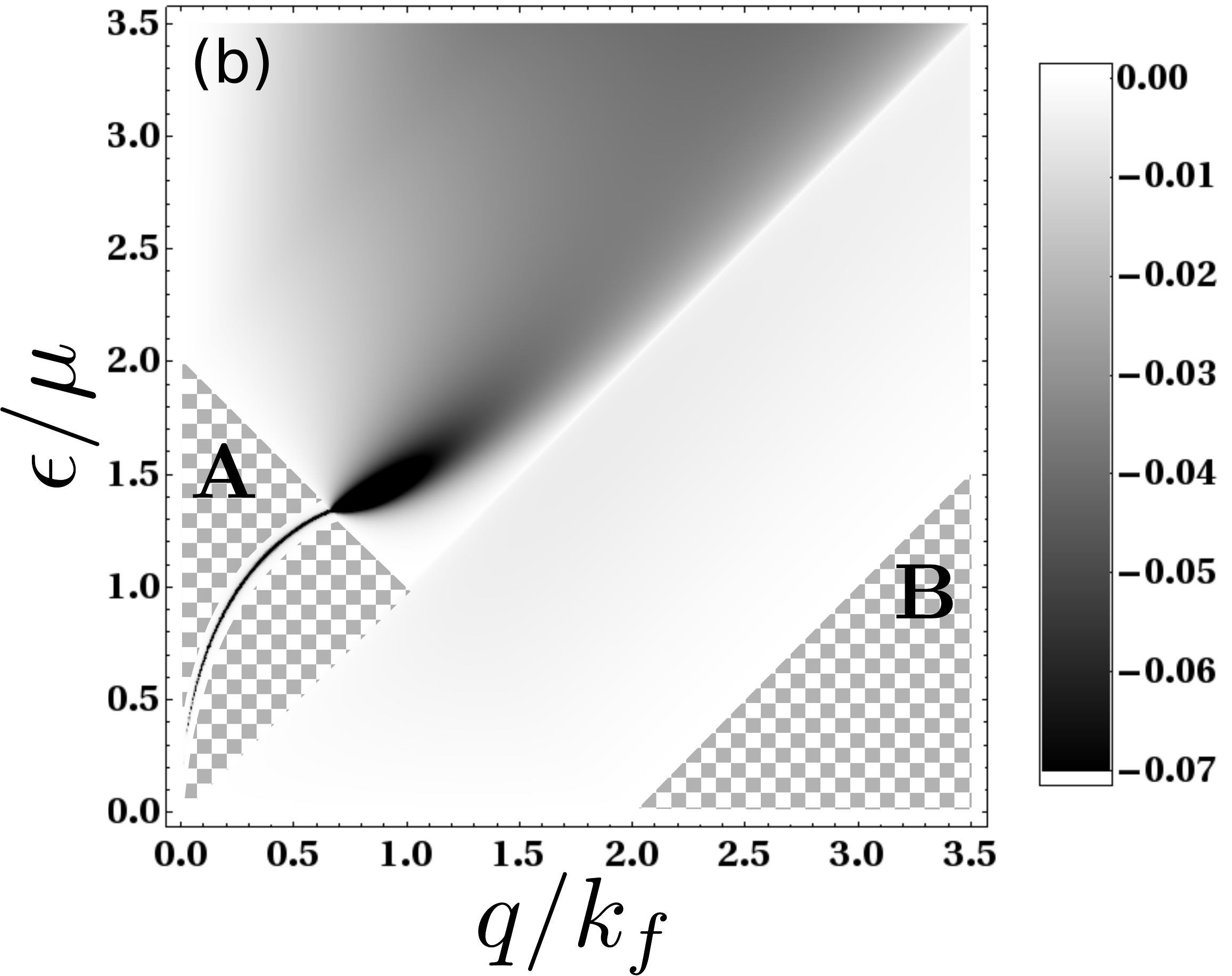}
\caption{\label{fig:P_Im} Density plot of (a) ${\rm Im}\left[\Pi^r_0(q,\epsilon)\right]$ and (b) ${\rm Im}\left[\Pi^r_{RPA}(q,\epsilon)\right]$ in units of $\frac{\mu}{\hbar^2v^2_f}$. Checkerboard pattern emphasizes regions where the imaginary part of the polarization operator vanishes exactly.}
\end{figure}

Polarization operator $\Pi^r(q,\epsilon)$ for graphene can be obtained at various levels of theory. At zeroth order approximation with respect to the electron-electron Coulomb interaction within the graphene sheet, it can be evaluated adopting the free massless Dirac fermions (MDF) approximation. At this level, the polarization operator, denoted by $\Pi^r_0(q,\epsilon)$, is a bare polarization bubble describing a single non-interacting electron-hole pair in graphene. Within this approximation, $\Pi^r_0(q,\epsilon)$ has been evaluated previously\cite{Shung1986-979,Wunsch2006-318,Hwang2007-205418} for arbitrary doping level and its imaginary part is shown in Fig.~\ref{fig:P_Im}(a).  
This figure is valid for any level of doping, defined by the chemical potential, $\mu$, measured relative to the Dirac point, since due to the linear dispersion relation of massless fermions in graphene, the polarization operator does not change with doping, if measured in units of $\frac{\mu}{\hbar^2 v^2_f}$ and plotted against unitless coordinates $q/k_f$ and $\epsilon/\mu$, where $v_f\approx$~0.4 is the Fermi velocity in atomic units and $k_f=\mu/\hbar v_f$.\cite{Peres2006-125411}

With respect to the wave number, $q$, the MDF approximation is only valid when $q\ll a^{-1}$, where $a\approx$~2.46\AA~is the lattice constant of graphene. Because of the factor $e^{-2qz}$ in the integrand of Eq.~(\ref{eq:QE}), this condition is equivalent to restricting $z\gg a$, i.e., to considering the quenching efficiency $\varphi_q$ only at large QD-graphene distances compared to the lattice constant of graphene. However, because of the adopted dipole approximation we are already restricted to $z\gg D$, where the typical QD diameter ($D$) is on the order of a few nanometers. Thus, setting $z\gg a$ because of the MDF approximation does not restrict the range of applicability of Eq.~(\ref{eq:QE}) any further compared to the one set by the dipole approximation. With respect to the excitation energy, the MDF approximation has been shown to be valid for $\epsilon\lesssim 1$~eV.\cite{Hill2009-27005}

The key features of ${\rm Im}\left[ \Pi^r_0(q,\epsilon)\right]$ include (i) the singularity along the $\epsilon/\mu=q/k_f$ line, which corresponds to single-particle excitations with $({\bf q},\epsilon)$ vector lying within the surface of the Dirac cone, and (ii) the absence of single-particle excitations, i.e., ${\rm Im}\left[ \Pi^r_0(q,\epsilon)\right]\equiv 0$, in regions $\bf A$ and $\bf B$, marked by checkerboard patterns in Fig.~\ref{fig:P_Im}(a). Equation~\ref{eq:qratef} with $\Pi^r_0(q,\epsilon)$ substituted in is very similar\cite{fn2} to the result of Swathi and Sebastian\cite{Swathi2008-054703,Swathi2009-086101} for the fluorescence quenching of a dye molecule due to single-particle excitations in graphene.

The polarization operator within the bare bubble approximation does not include the graphene's polarization self-consistently which can become crucial at  non-zero doping levels ($\mu>0$), where the finite carrier density at the Fermi level leads to the efficient Coulomb screening within the graphene's sheet. To correct for this, we evaluate the polarization operator within the random-phase approximation (RPA) as\cite{Wunsch2006-318,Hwang2007-205418}
\begin{equation}
\Pi^r_{RPA}(q,\epsilon)=\frac{\Pi^r_0(q,\epsilon)}{1-W(q)\Pi^r_0(q,\epsilon)},
\label{eq:PRPA}
\end{equation}
where $W(q)=2\pi e^2/\tilde{\kappa} q$ is the two-dimensional Fourier transform of the Coulomb potential within the graphene's plane. The effective dielectric constant of the environment is denoted by $\tilde{\kappa}$. A free-standing graphene sheet in vacuum corresponds to $\tilde{\kappa}=1$. For graphene laying on top of a half-space dielectric substrate with dielectric constant $\kappa$, the effective constant is given by $\tilde{\kappa}=(\kappa+1)/2$.\cite{Smythe1968,Ponomarenko2009-206603} For example, for a SiO$_2$ substrate ($\kappa$=4) the effective dielectric constant is $\tilde{\kappa}=2.5$. In the rest of the paper, except for results shown in Fig.~\ref{fig:disps}, the vacuum conditions ($\tilde{\kappa}=1$) are assumed.

The imaginary part of $\Pi^r_{RPA}(q,\epsilon)$ is depicted in Fig.~\ref{fig:P_Im}(b). It is seen that the ``single-particle'' singularity $\epsilon/\mu=q/k_f$ is gone and the new singularity at $\epsilon\propto q^{1/2}$ appears instead. This emergent singularity corresponds to the collective electronic excitation, i.e., plasmon, in graphene.\cite{Shung1986-979,Wunsch2006-318,Hwang2007-205418} It is rather pedagogical to see exactly how this singularity appears from Eq.~(\ref{eq:PRPA}). The naive substitution of $\Pi^r_{0}(q,\epsilon)$, shown in Fig.~\ref{fig:P_Im}, into Eq.~(\ref{eq:PRPA}) results in  ${\rm Im}\left[\Pi^r_{RPA}(q,\epsilon)\right]$ vanishing exactly in the entire region $\bf A$ since both $\Pi^r_{0}(q,\epsilon)$ and $W(q)$ are real in this region. However, the more careful analysis reveals that the imaginary part of $\Pi^r_{0}(q,\epsilon)$ is not exactly zero in $\bf A$, but infinitesimal instead due to the usual small imaginary constant in the denominator of the Lindhard function, which preserves causality. The infinitesimal imaginary part of the bare polarization operator can be safely neglected if $\Pi^r_{0}(q,\epsilon)$ is analyzed by itself. However, when $1-W(q){\rm Re}\left[\Pi^r_{0}(q,\epsilon)\right]$ vanishes in the denominator of Eq.~(\ref{eq:PRPA}), this small imaginary part yields singularity ($\delta$ function) in ${\rm Im}\left[\Pi^r_{RPA}(q,\epsilon)\right]$, i.e.,
\begin{equation}
{\rm Im}\left[\Pi^r_{RPA}(q,\epsilon)\right]\propto \delta(q-q_p(\epsilon)),
\label{eq:RPA_delta}
\end{equation}
where the plasmon dispersion relation is given by $q_p(\epsilon)\approx\frac{\epsilon^2}{2e^2\mu}$ at small $\epsilon$, i.e., at $\epsilon/\mu\ll 1$, and obtained numerically at higher excitation energies.\cite{Hwang2007-205418} Therefore, the general prescription to assume all the infinitesimal parameters finite till the very end of calculations proves to be critical in this case.

It is important to note that since there is no single-particle excitations in region $\bf A$, the plasmon lifetime is infinite [$\delta$ function in the imaginary part of $\Pi^r_{RPA}(q,\epsilon)$] within this region.\cite{fn3} Once plasmon ``leaves'' region $\bf A$ ($q/k_f\gtrsim 0.65$), it acquires the finite lifetime due to the Landau damping and soon disappears.

In the following two sections, \ref{sec:sp} and \ref{sec:plasmon}, we discuss the efficiency of the QD fluorescence quenching due to unscreened excitations in graphene, described by $\Pi^r_0(q,\epsilon)$, as well as screened excitations in graphene, described by $\Pi^r_{RPA}(q,\epsilon)$, respectively.

\section{Quenching by unscreened excitations}
\label{sec:sp}

QE due to unscreened single-particle excitations in graphene can be obtained from Eq.~(\ref{eq:QE}) by substituting $\Pi^r(q,\epsilon)$ with $\Pi^r_0(q,\epsilon)$. This approximation is accurate if either (i) the effective dielectric constant of a substrate is high, and, therefore, the effective electron-electron Coulomb interaction within graphene is greatly reduced, or (ii) $\epsilon\gg \mu$, i.e., graphene becomes effectively undoped. In both cases, the screening within the graphene's layer becomes ineffective yielding $\Pi^r_{RPA}(q,\epsilon)\approx \Pi^r_0(q,\epsilon)$.

\begin{figure}
\centering{}
\vspace{0.2in}
\includegraphics[width=7cm]{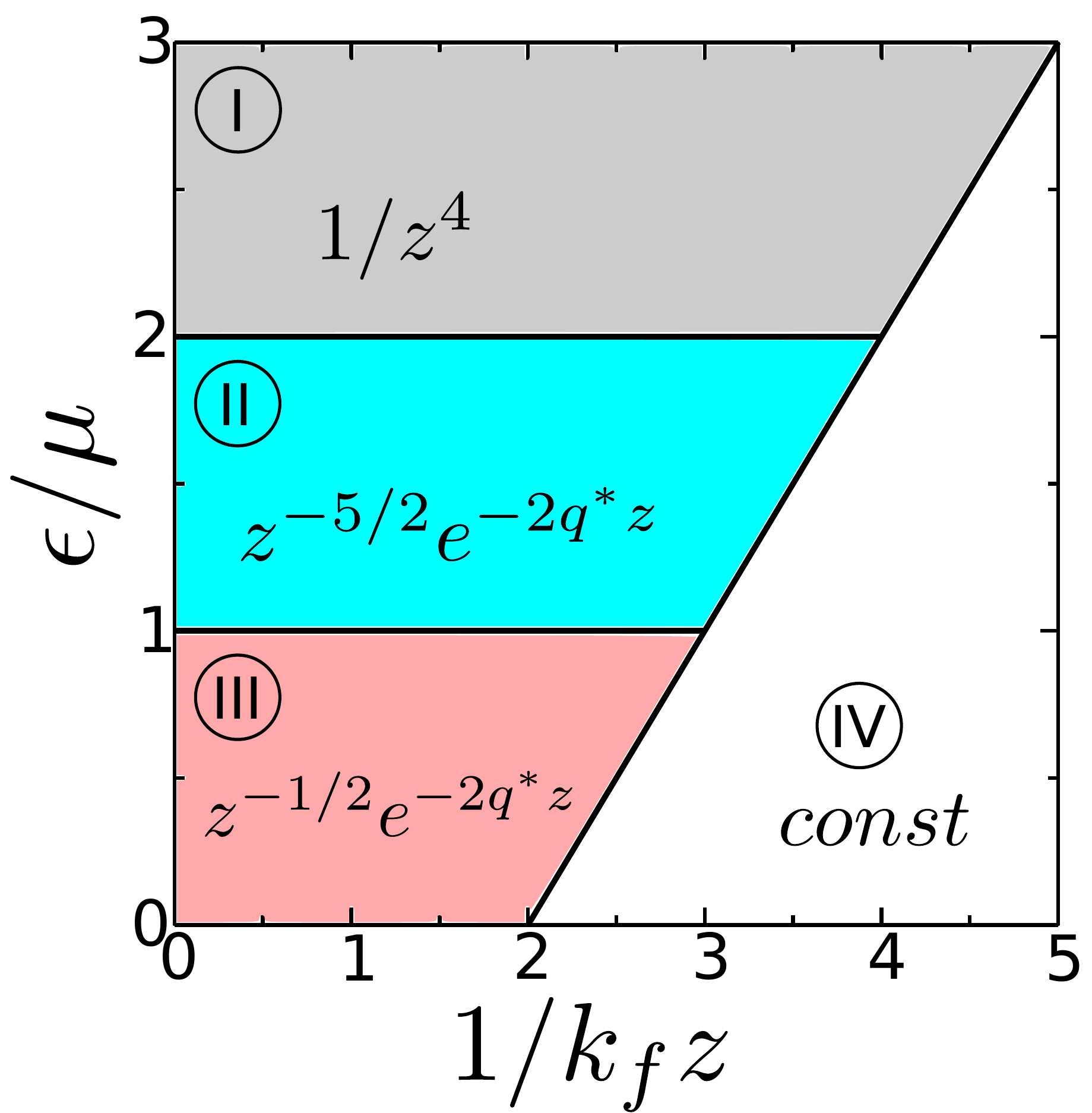} \caption{\label{fig:regimes} Regimes of asymptotic dependence of quenching efficiency on the QD-graphene distance, $z$, at $z\rightarrow\infty$ (Regimes I, II and III) and at $z\rightarrow 0$ (Regime IV).}
\end{figure}

Four different regimes of the asymptotic dependence of QE on the QD-graphene distance, $z$, are shown schematically in Fig.~\ref{fig:regimes}.
At large distances, exponent $e^{-2qz}$ in the integrand of Eq.~(\ref{eq:QE}) decays rapidly with $q$, which guarantees that at fixed excitation energy $\epsilon$ the dominant contribution to QE comes from lowest possible $q$'s where the imaginary part of the polarization operator is still nonzero. The two qualitatively different cases are (i) ``gapless'', where the imaginary part of the polarization operator is already non-zero even at infinitesimally small $q>0$, and (ii) ``finite-gap'', where ${\rm Im}\left[\Pi^r_{0}(q,\epsilon)\right]$ becomes non-zero only at some finite $q$, i.e., at $q>q^*$ with finite $q^*>0$. The typical power law dependence of $\Pi^r_0(q,\epsilon)$ on $q$ at $q\rightarrow 0$ in the gapless case
\begin{subequations}
\begin{equation}
\Pi^r_0(q,\epsilon)\propto q^\alpha,
\end{equation}
and 
\begin{equation}
\Pi^r_0(q,\epsilon)\propto(q-q^*)^\alpha,
\end{equation}
\end{subequations}
at $q\rightarrow  q^*+0$ in the finite-gap case, yield
\begin{subequations}
\begin{equation}
\varphi_q\propto z^{-(\alpha+2)},
\label{eq:gapless_asymp}
\end{equation}
and
\begin{equation}
\varphi_q\propto z^{-(\alpha+1)}e^{-2q^*z},
\label{eq:gap_asymp}
\end{equation}
\end{subequations}
at $z\rightarrow\infty$, respectively. In what follows, we consider three different asymptotic regimes of QE at $z\rightarrow \infty$ (Regimes I, II and III). The asymptotic behavior of QE at $z\rightarrow 0$ (Regime IV) is described subsequently.

\paragraph*{Regime I.}
At high excitation energies ($\epsilon/\mu>2$) the imaginary part of $\Pi^r_0(q,\epsilon)$ is seen in Fig.~\ref{fig:P_Im}(a) to correspond to the gapless case. Further, it can be shown that ${\rm Im}\left[\Pi^r_0(q,\epsilon)\right]\propto q^2$ at small $q$ (see Eq.~(12) in Ref.~\onlinecite{Wunsch2006-318} for the long wavelength limit of the polarization operator), which is then combined with Eq.~(\ref{eq:gapless_asymp}) to give
\begin{equation}
\varphi_q\propto 1/z^4.
\label{eq:z-4}
\end{equation}
This power-law dependence is shown  as Regime I in Fig.~\ref{fig:regimes}. The asymptotic behavior in Eq.~(\ref{eq:z-4}) is satisfied at $\epsilon/\mu>2$, and, therefore, also at $\epsilon\gg\mu$, where $\mu$ can be treated as zero, i.e., the graphene sheet becomes effectively undoped. The asymptotics of $1/z^4$ for the fluorescence quenching by single-particle excitations in undoped (and also weakly doped) graphene was first predicted by Swathi and Sebastian.\cite{Swathi2008-054703,Swathi2009-086101} 

Generally, the asymptotic dependence $1/z^4$ is rather expected since QE due to the energy transfer from a chromophore to a quencher accompanied by the interband excitation of single electron-hole pairs in an $N$-dimensional quencher typically scales as $z^{N-6}$, where $z$ is the distance between the chromophore and the quencher. FRET to $0d$ quencher (e.g., small organic molecule) corresponds to $N=0$, naturally resulting in $1/z^{6}$.\cite{Forster1965} Nanowire as a quencher ($N=1$) leads to $1/z^5$ asymptotics.\cite{Hernandez-Martinez2008-035314} Excitation of single-electron holes in a dipole-to-surface configuration, i.e., when a quencher occupies the half-space ($N=3$), yields $1/z^3$ asymptotics.\cite{Persson1982-5409} The graphene sheet is a $2d$ object, which naturally leads to $1/z^4$ dependence of QE at large $z$.

\paragraph*{Regime II.}
At $1<\epsilon/\mu<2$, the imaginary part of $\Pi^r_0(q,\epsilon)$ vanishes exactly at $q<q^*$, where the threshold value of $q^*/k_f=2-\epsilon/\mu$ marks the onset of {\em interband} single-particle excitations. It can be shown that in this finite-gap case, ${\rm Im}\left[\Pi^r_0(q,\epsilon)\right]$ is proportional to $(q-q^*)^{3/2}$ at $q\rightarrow q^*+0$, which yields [by virtue of Eq.~(\ref{eq:gap_asymp})]
\begin{equation}
\varphi_q\propto z^{-5/2}e^{-2q^* z},
\end{equation}
which is marked as regime II in Fig.~\ref{fig:regimes}.

\paragraph*{Regime III.}
At low excitation energies, $\epsilon/\mu<1$, the finite-gap case is again realized with $q^*/k_f=\epsilon/\mu$, which is the onset of {\em intraband} single-particle excitations. In this regime, ${\rm Im}\left[\Pi^r_0(q,\epsilon)\right]\propto (q-q^*)^{-1/2}$, resulting in
\begin{equation}
\varphi_q\propto z^{-1/2}e^{-2q^* z},
\end{equation}
depicted as regime III in Fig.~\ref{fig:regimes}.

\paragraph*{Regime IV.}
Finally, at a fixed excitation energy there are no single-particle excitations with $q/k_f>2+\epsilon/\mu$, i.e, in region $\bf B$ in Fig.~\ref{fig:P_Im}. This introduces the natural high-$q$ cutoff for integration in Eq.~(\ref{eq:QE}). If, at certain (small) $z$, exponent $e^{-2qz}$ is still $\approx 1$ at $q$ approaching this cutoff, then QE becomes constant with respect to $z$. This is depicted as regime IV in Fig.~\ref{fig:regimes}. However, we expect this regime to be hardly accessible experimentally. For example, at $\epsilon=0.8$~eV and $\mu=0$, the realization of this regime requires $z<$5~\AA. At such small distances photoinduced charge transfer from QD to graphene can become the dominant mechanism of fluorescence quenching  prohibiting the analysis of the relatively less efficient energy transfer channel.\cite{RamakrishnaMatte2011} Besides, at such distances both dipole and MDF approximations are likely to break down.

\section{Quenching by screened excitations}
\label{sec:plasmon}

To evaluate QE in the case of screened excitations in graphene, one has to substitute $\Pi^r(q,\epsilon)$ in Eq.~(\ref{eq:QE}) with $\Pi^r_{RPA}(q,\epsilon)$ given by Eq.~(\ref{eq:PRPA}). First, we consider regime I in Fig.~\ref{fig:regimes}. In this regime, the product $W(q)\Pi^r_0(q,\epsilon)$ becomes proportional to $1/q\times q^2=q$ at $q\rightarrow 0$, which results in the approximate equality $\Pi^r_{RPA}(q,\epsilon)\approx \Pi^r_0(q,\epsilon)$ at small $q$. Thus, the asymptotic behavior of $\varphi_q(z)$ at large $z$ is the same for screened and unscreened excitations in regime I; i.e., taking screening into account does not lead to qualitative changes in QE. This can be easily understood for $\epsilon\gg\mu$, where graphene becomes effectively undoped, and, therefore, the small free carrier density renders screening within graphene inefficient.  

The situation is different in regime III ($\epsilon/\mu<1$), where taking Coulomb screening into account within RPA leads to the emergence of the new (plasmon) singularity in the polarization operator, as is shown in Fig.~\ref{fig:P_Im}(b), region $\bf A$. Equation~(\ref{eq:RPA_delta}) implies that ${\rm Im}\left[\Pi^r_{RPA}(q,\epsilon)\right]$ vanishes exactly at $q<q^*=q_p(\epsilon)$; i.e., the finite-gap situation is realized with $q^*$ defined by the plasmon dispersion relation. Substituting Eq.~(\ref{eq:RPA_delta}) into Eq.~(\ref{eq:QE}), one obtains
\begin{equation}
\varphi_q(z)\propto e^{-2q_p(\epsilon) z}.
\label{eq:Asym_plasm}
\end{equation}
The absence of the power-law multiplier in front of the exponent, which was universally present in the finite-gap situations in the previous section, is related to a delta-functional instead of a power-law singularity of the polarization operator.
 
The asymptotic behavior, given by Eq.~(\ref{eq:Asym_plasm}), is correct even outside regime III, since the plasmon branch in Fig.~\ref{fig:P_Im}(b) remains singular up to $\epsilon/\mu\approx 1.3$, i.e., well within what used to be regime II in the case of unscreened excitations. For $1.3\lesssim\epsilon/\mu<2$ the imaginary part of $\Pi^r_{RPA}(q,\epsilon)$ scales as $(q-q^*)^{3/2}$ at $q\rightarrow q^*+0$ with the $q$ gap defined by $q^*/k_f=2-\epsilon/\mu$, which yields $\varphi_q(z)\propto z^{-5/2}e^{-2q^* z}$, i.e., the large-$z$ asymptotics is identical to that of regime II in the case of unscreened excitations.

To examine how accurately large-$z$ asymptotics for $\varphi_q(z)$ reproduce exact solutions at finite $z$, we numerically evaluate the integral in Eq.~(\ref{eq:QE}) for the realistic case of PbSe QD with the excitation energy of $\epsilon=0.8$~eV. At this energy, the fluorescence quantum yield of the isolated PbSe QD can be as high as 70--90\% with fluorescence lifetimes up to 1 $\mu$s,\cite{Du2002-1321,Liu2010-14860} suggesting that QE can be directly extracted from experiment, since the observable fluorescence quantum yield is given by $1/(1+\varphi_q)$ in the presence of FRET.\cite{fn4}

\begin{figure}
\centering{}
\vspace{0.2in}
\includegraphics[width=7.5cm]{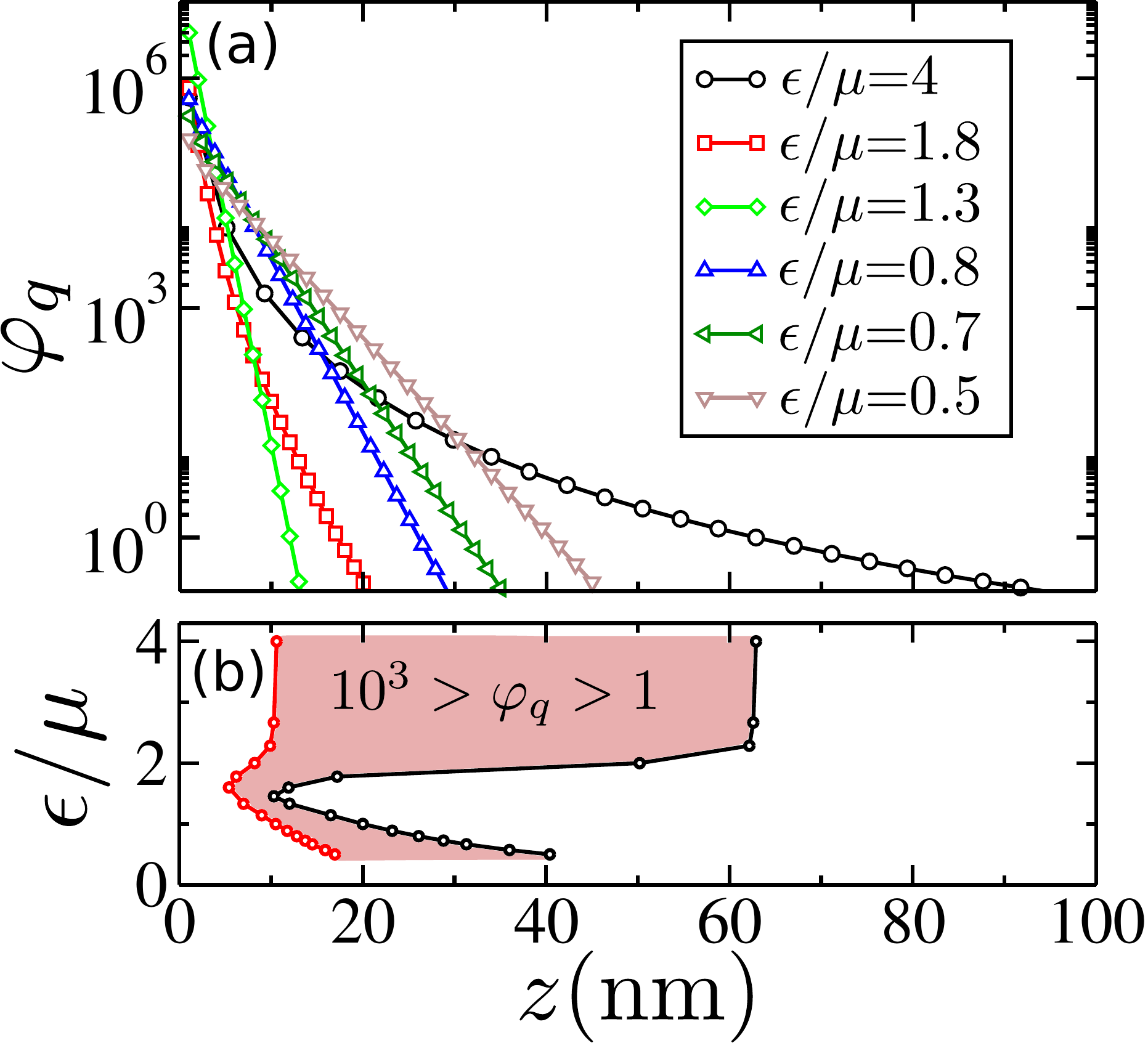} \caption{\label{fig:zdeps} (a) Dependence of the quenching efficiency $\varphi_q$ on the QD-graphene distance $z$. (b) The filled area shows where $\varphi_q$(z) is between 1000 and 1 at a given excitation energy $\epsilon/\mu$. The red (left) and black (right) circles mark $\varphi_q(z)$=1000 and $\varphi_q(z)$=1 contour lines, respectively.}
\end{figure}

Numerically evaluated $\varphi_q(z)$ for several values of chemical potential in the range $\mu=0.2-1.6$ eV is shown in Fig.~\ref{fig:zdeps}(a).
The smallest value of chemical potential adopted corresponds to Regime I since $\epsilon/\mu=4$. The corresponding $\varphi_q(z)$, depicted by black circles, is expected to show $1/z^4$ dependence at large $z$, and, indeed, demonstrates slowly decaying non-exponential tail.

All the other values of chemical potential give $\epsilon/\mu<2$, and, therefore, are expected to give $\varphi_q(z)\propto e^{-2q_p(\epsilon) z}$ and $\varphi_q(z)\propto z^{-5/2}e^{-2q^* z}$ at large $z$ for $\epsilon/\mu\lesssim1.3$ and $1.3\lesssim\epsilon/\mu<2$, respectively, with $q^*$ defined by $q^*/k_f=2-\epsilon/\mu$. As expected, all $\varphi_q(z)$'s, except for the one corresponding to the lowest value of chemical potential ($\mu=0.2$~eV), demonstrate the nearly exponential decay at large $z$. The filled area in Fig.~\ref{fig:zdeps}(b) shows the range of QD-graphene distances, where QE is between 1 and 1000 -- somewhat loosely chosen range where the accurate experimental measurement of QE is still possible. QE is seen to decay the fastest with $z$ (lowest $z$ at fixed $\varphi_q$) where the $q$-gap is the largest, i.e., at $\epsilon/\mu\approx 1.3$, as is seen in Fig.~\ref{fig:P_Im}(b). 

\begin{figure}
\centering{}
\vspace{0.2in}
\includegraphics[width=8cm]{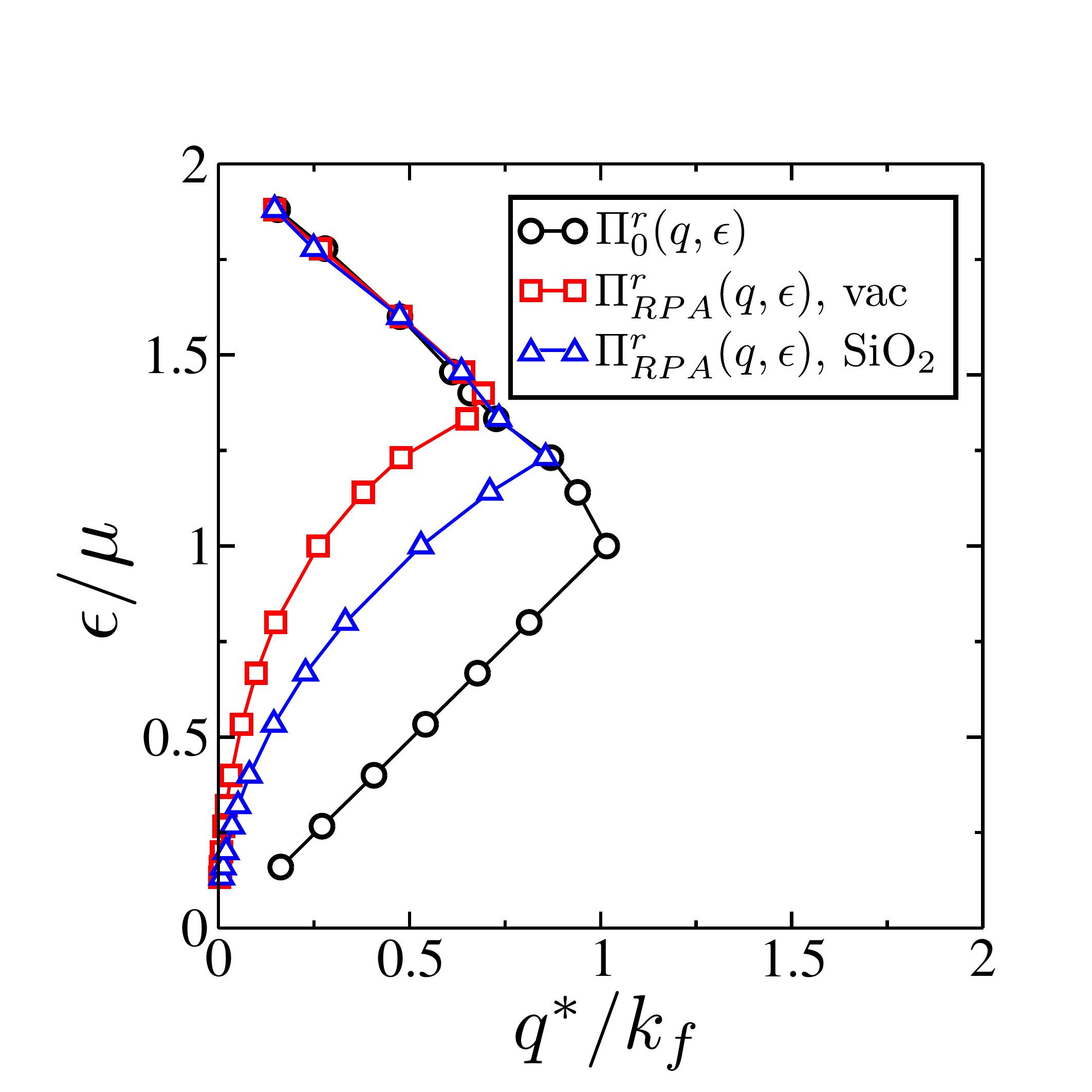} \caption{\label{fig:disps} QE decay rate $q^*$ plotted vs $\epsilon/\mu$. Circles (black), squares (red) and triangles (blue) correspond to single-particle approximation, RPA for free-standing graphene (vacuum) and RPA for graphene on substrate (SiO$_2$, $\kappa$=4), respectively.}
\end{figure}

The rate of the exponential decay of QE at $\epsilon/\mu<2$ directly reflects the specific structure of ${\rm Im}\left[\Pi^r_{RPA}(q,\epsilon)\right]$ at small $q$, i.e., the width of the finite $q$-gap. Therefore, the large-$z$ behavior of QE can be used to extract the valuable information about electronic excitations in graphene, e.g., the plasmon dispersion. To illustrate how the plasmon dispersion can be extracted, we plot $q^*$ against $\epsilon/\mu$ in Fig.~\ref{fig:disps}, where $q^*$ is the decay rate of QE at large $z$, obtained by fitting the large-$z$ QE decay with $\varphi_q(z)\propto e^{-2q^* z}$. 
Specifically, $q^*$ is plotted for the case of unscreened excitations (black circles), as well as for screened excitations in free-standing graphene (red squares) and graphene laying on top of the SiO$_2$ substrate (blue triangles). Scanning through the range of values of chemical potential using the back-gate, we are able to extract the dispersion relation of singularities of the imaginary part of the polarization operator. As is seen, the linear dispersion relation of single-particle excitations (black circles at $\epsilon/\mu<1$) as well as that of the plasmon in the free-standing graphene (red squares at $\epsilon/\mu\lesssim 1.3$) in graphene is accurately recovered.

As discussed in the beginning of Sec.~\ref{sec:sp}, electronic excitations in graphene are effectively single-particle either if (i) there is a strong substrate-induced dielectric screening or (ii) the excitation energy is high, i.e., $\epsilon\gg\mu$. The first case is illustrated by the plasmon dispersion for graphene on the SiO$_2$ substrate (blue triangles), which is closer to the single-particle excitations (black circles) than that corresponding to the free-standing graphene (red squares). The second case is effectively realized at $\epsilon/\mu\gtrsim 1.3$, where dispersions of electronic excitations with and without accounting for the in-graphene screening are nearly identical.

The proposed method to extract the plasmon dispersion requires an accurate control over the distance between PbSe QD and the graphene sheet. This can be accomplished by either using the core-shell type-1 structures\cite{Reiss2009-154} or by growing a dielectric layer on top of graphene by atomic layer deposition with the controllable thickness and then depositing QDs on top of that dielectric layer. The first approach can be based on PbSe/CdSe core-shell structures, where the large bulk bandgap of CdSe ($\sim$1.7 eV) as compared to that of bulk PbSe ($\sim$0.28 eV) guarantees the confinement of both electron and hole within the PbSe core (type-1 structure).\cite{Pietryga2008-4879} This confinement guarantees that the shell serves only as an inert spacer between the PbSe core and the graphene layer. Recent advances in core-shell structure fabrication techniques allow one to control the thickness of the shell with the monolayer (subnanometer) precision.\cite{Reiss2009-154} Atomic layer deposition provides an alternative strategy for controlling the distance between QD and graphene with sub-nanometer resolution.\cite{Alaboson2011-5223}

Finally, we wish to discuss the effect of the finite plasmon propagation length on the applicability of the proposed method. So far, we assumed that the plasmon has infinite propagation length within region ${\bf A}$ in Fig.~\ref{fig:P_Im}(b). However, additional damping channels, not accounted for in RPA (e.g., defect scattering, electron-phonon coupling), can ``broaden'' the plasmon in region ${\bf A}$. It follows from Eq.~(\ref{eq:QE}), that if the plasmon width in the $q$ domain, $\delta q$, is less then $1/z$, then ${\rm Im}\left[\Pi^r_{RPA}(q,\epsilon)\right]$ can still be treated as $\delta$ function, resulting in Eq.~(\ref{eq:Asym_plasm}). Accordingly, deviations from the exponential dependence are expected  at $z\gtrsim\delta q^{-1}\sim l$, where $l$ is the plasmon propagation length.

Our numerical tests show that the exponential decay of QE is typically already established at $z$ comparable with the plasmon wavelength $\lambda$. On the other hand, estimations by various authors suggest that the plasmon propagation length could be as high as 10-100 $\lambda$.\cite{Jablan2009-245435,Koppens2011-arXiv1104.2068} Therefore, we expect QE to decay exponentially at $z\gtrsim\lambda$, thus allowing for extraction of plasmon dispersion. The onset of deviations from this exponential decay at larger QD-graphene distances, $z\gtrsim l$, naturally provides a good estimate for the plasmon propagation length.

\section{Conclusion}
\label{sec:conclusion}

Based on the detailed analysis of fluorescence quenching efficiency in the QD-graphene complex, we have proposed a method of probing and studying electronic excitations in graphene. The method has been demonstrated to be sufficiently sensitive to allow the extraction of the dispersion relation of plasmon in graphene. We hope that this study will stimulate experimental efforts in this direction, especially because the proposed method is based on the QD-graphene complex which can be of interest not only as a means to probe electronic excitations in graphene, but also on its own merit as a key component of hybrid nanostructures with promising properties. For example, the ability to excite plasmon locally in graphene using a semiconductor quantum dot can become of great use in graphene plasmonics. Another recently proposed use of QD-graphene complexes is in photovoltaics.\cite{Sun2011-093112}

The RPA-based description of electronic excitations in graphene, adopted in this paper, does not include such many-body effects as exchange and correlation, impurity and defect scattering, or electron-phonon coupling. These effects can affect the plasmonic response of graphene in two ways. First, there can be deviations of the actual plasmon dispersion relation from the one shown in Fig.~\ref{fig:P_Im}(b). Second, additional channels of plasmon damping can appear, as discussed in the previous section. The proposed experimental technique is capable of assessing both the dispersion relation and the damping and, thus, is expected to stimulate the development of beyond-RPA theoretical methods by providing a necessary experimental validation. 

\acknowledgments

This work was performed, in part, at the Center for Integrated Nanotechnologies, a U.S. Department of Energy, Office of Basic Energy Sciences user facility. K.A.V. acknowledges support by the Center for Nonlinear Studies (CNLS), LANL.

\appendix*

\section{Derivation of quenching rate}
\label{app:qrate}

Evaluating explicitly the QD part of the matrix element in Eq.~(\ref{eq:qrate}) one obtains 
\begin{equation}
k_q=2\pi\hbar^{-1}\sum_n\left|\langle n|\int_g d{\bf r}\:V({\bf r})\hat{\rho}({\bf r})|0\rangle\right|^2\delta(\epsilon-E_n).
\end{equation}
This can be rewritten as
\begin{align}
k_q&=\frac{\hbar^{-1}}{(2\pi)^3}\sum_n\int d{\bf r}d{\bf r}'\int d{\bf q}d{\bf q}'\:V^*({\bf q})e^{-i{\bf q}\cdot{\bf r}}V({\bf q}')e^{i{\bf q}'\cdot{\bf r}'}
\nonumber\\
&\times\rho_{0n}({\bf r})\rho_{n0}({\bf r}')\delta(\epsilon-E_n),
\end{align}
where $\rho_{n0}({\bf r})=\langle n|\hat{\rho}({\bf r})|0\rangle$ and $V({\bf r})=\frac{1}{(2\pi)^2}\int d{\bf q}\:V({\bf q})e^{i{\bf q}\cdot{\bf r}}$. Further, the $\delta$ function can be substituted using the identity $\delta(\epsilon-E_n)=\frac{\hbar^{-1}}{2\pi}\int dt\:e^{i(\epsilon-E_n)t/\hbar}$ yielding
\begin{align}
k_q&=\frac{i\hbar^{-1}}{(2\pi)^4}\int d{\bf r}d{\bf r}'\int d{\bf q}d{\bf q}'\:V^*({\bf q})e^{-i{\bf q}\cdot{\bf r}}V({\bf q}')e^{i{\bf q}'\cdot{\bf r}'}
\nonumber \\
&\times\int dt\:\Pi^>({\bf r},{\bf r}';t)e^{i\epsilon t/\hbar},
\end{align}
where $\Pi^>({\bf r},{\bf r}';t)=-i\hbar^{-1}\sum_n\rho_{0n}({\bf r})\rho_{n0}({\bf r}')e^{-iE_n t/\hbar}$ is the ``greater'' polarization operator for graphene in the Lehmann representation. At sufficiently large distances between the quantum dot and the graphene layer, $V({\bf r})$ varies smoothly within the graphene's plane. This makes it possible to average the polarization operator over the unit cell with respect to both ${\bf r}$ and ${\bf r}'$. After this averaging, the polarization operator becomes insensitive to variations of electronic density on the scale of the graphene's unit cell and, therefore, acquires the isotropy and the continuous translational symmetry instead of the discrete one, leading to a possibility to substitute $\Pi^>({\bf r},{\bf r}';t)\rightarrow \Pi^>(|{\bf r}-{\bf r}'|,t)$. Then, integrations over ${\bf r}$, ${\bf r}'$, and $t$ can be interpreted as spatial and time Fourier transforms, respectively, resulting in
\begin{equation}
k_q=\frac{i\hbar^{-1}}{(2\pi)^2}\int d{\bf q}\:\left|V({\bf q})\right|^2 \Pi^>(q,\epsilon),
\end{equation}
where
\begin{equation}
\Pi^>(q,\epsilon)=\int d{\bf r}\int dt\:\Pi^>(|{\bf r}|,t)e^{-i({\bf q}{\bf r}-\epsilon t/\hbar)}.
\end{equation}
Using the relations between real-time correlation and response functions at equilibrium,\cite{Abrikosov1963} which at zero temperature yields $\Pi^>=2i{\rm Im}[\Pi^r]$, we obtain
\begin{equation}
k_q=-\frac{2\hbar^{-1}}{(2\pi)^2}\int d{\bf q}\:\left|V({\bf q})\right|^2 {\rm Im}\left[\Pi^r(q,\epsilon)\right].
\label{eq:qrate1}
\end{equation}
This equation is similar to the well-known result by Metiu [Eq.~(2.30) in Ref.~\onlinecite{Metiu1984-153}].

Finally, we provide without derivation the two-dimensional Fourier transform (within the graphene's plane) of $V({\bf r})$
\begin{equation}
V({\bf q})=2\pi i e(d_\parallel\cos(\theta)+i d_\perp)e^{-qz},
\end{equation}
where ${\bf d}_\parallel$ is the projection of the QD transition dipole onto the graphene's plane, $d_\parallel=|{\bf d}_\parallel|$ being its magnitude. The angle between vectors ${\bf d}_\parallel$ and ${\bf q}$ is denoted by $\theta$. The projection of the QD transition dipole onto the normal to the graphene's plane is denoted by $d_\perp$. Substitution of $V({\bf q})$ into Eq.~(\ref{eq:qrate1}) and the subsequent integration over $\theta$ yields Eq.~(\ref{eq:qratef}).


\begin{thebibliography}{43}
\bibitem[{fn1()}]{fn1}
\bibinfo{note}{Here, the term ``chromophore'' is used rather loosely referring
  to an object (e.g., a molecule, semiconductor quantum dot), whose electronic
  degrees of freedom interact with the external electromagnetic field.}
\bibitem[{fn3()}]{fn3}
\bibinfo{note}{The lack of single-particle excitations leads only to the
  absence of Landau damping within the RPA-based description. Other mechanisms
  of plasmon damping, e.g., due to excitation of phonons or multiple
  electron-hole pairs, can still be present.}
\bibitem[{fn4()}]{fn4}
\bibinfo{note}{This expression is only accurate if the fluorescence quantum
  yield of the isolated QD approaches 100\%. Otherwise, it has to be corrected
  to include the nonradiative deexcitation of the isolated QD.}
\end{thebibliography}

\end{document}